\documentclass[12pt,letterpaper]{article}

\usepackage{setspace}
\usepackage{geometry}
\usepackage{float}
\usepackage{graphicx}
\usepackage{endfloat}
\usepackage{amssymb}
\usepackage{amsmath}
\usepackage{natbib}

\usepackage{tikz}
\usetikzlibrary{positioning}
\usepackage{lscape}
\usepackage{bm}

\usepackage[normalem]{ulem}

\newcommand{\bbeta}{\boldsymbol{\beta}}

\newcommand{\T}{\top}
\newcommand{\bS}{\mathbf{S}}

\newcommand{\bQ}{\mathbf{Q}}

\newcommand{\bX}{\mathbf{X}}
\newcommand{\bP}{\mathbf{P}}
\newcommand{\bSigma}{\boldsymbol{\Sigma}}
\newcommand{\bee}{\mathbf{e}}

\begin{document}

\title{Understanding the stochastic partial differential equation approach to smoothing}

\author{David L Miller \thanks{Joint first author. \textit{Address:} Centre for Research into Ecological \& Environmental Modelling and School of Mathematics \& Statistics, University of St Andrews, St Andrews, Fife, Scotland} \and Richard Glennie\footnotemark[1] \and Andrew E Seaton\footnotemark[1]}

\date{} 

\maketitle

\begin{abstract}
  Correlation and smoothness are terms used to describe a wide variety of random quantities. In time, space, and many other domains, they both imply the same idea: quantities that occur closer together are more similar than those further apart. Two popular statistical models that represent this idea are basis-penalty smoothers \citep{wood_generalized_2017} and stochastic partial differential equations (SPDE) \citep{lindgren_explicit_2011}. In this paper, we discuss how the SPDE can be interpreted as a smoothing penalty and can be fitted using the \texttt{R} package \texttt{mgcv}, allowing practitioners with existing knowledge of smoothing penalties to better understand the implementation and theory behind the SPDE approach. 
\end{abstract}


\section{Introduction}
\label{s:introduction}

Data collected over space or time are often obtained with the desire to elicit an underlying pattern. The stochastic partial differential equation (SPDE) approach introduced by \citet{lindgren_explicit_2011} and implemented in the \texttt{R-INLA} software package \citep{rue_approximate_2009} is a flexible, efficient method to analyse such data. Despite this, wider application is inhibited by two obstacles. First, the methods are presented using mathematical concepts and terms more usually found in applied mathematics and physics, making it difficult for practitioners in other fields to understand and adapt these methods to their own needs. Second, available software implementations are difficult to customise without high-level technical knowledge, limiting application to only those models available in the software or specially requested from software developers. 

Here, we aim to mitigate these two issues: we describe ($i$) how the SPDE model can be interpreted as a basis-penalty smoother, a modelling framework more familiar to practitioners who use smoothing techniques \citep{wood_generalized_2017}, and ($ii$) how software to fit these smoothers (e.g., \texttt{mgcv}), regularly extended and customised for application, can be used to fit SPDE models or, to go further, used to incorporate SPDE methods into larger models. 

In this paper, we consider the following situation. Let \(z(x)\) be a random variable observed at location \(x\) or time \(x\), depending on the domain. A statistical model for \(z\) is constructed in three components or terms: \(z(x) = \eta(x) + f(x) + \epsilon(x)\). The first component, \(\eta(x)\), is the fixed effect, often a linear combination of observed covariates with unknown parameters. The third component, \(\epsilon(x)\), represents the measurement error or unstructured error, often \(\epsilon(x) \sim \mathcal{N}(0, \sigma^2)\) for unknown parameter \(\sigma\) and every location \(x\).
The second component is a stochastic process, representing the structured dependence among observations: observations made closer together in time or space are more likely to be similar than those further apart.  A mathematically convenient and flexible model for this component is a Gaussian process (GP) with mean zero and covariance function \(c(x_i, x_j) = \text{Cov}\left\{f(x_i), f(x_j)\right\}\). The covariance function quantifies how related two values of \(f\) are at two locations. For 
fixed locations \(\{x_1, \ldots, x_n\}\), the value of the GP at these locations, \(\{f(x_1), \ldots, f(x_n)\}\), are multivariate Gaussian with mean zero and covariance matrix \(\bSigma\) with \((i,j)^{\text{th}}\) entry \(c(x_i, x_j)\). We can extend this formulation to non-Gaussian responses by using a link function, $g$, so the response is then modelled with a specified distribution and a mean \(g^{-1}(\eta(x) + f(x))\).

\begin{figure}[H]
\centering
\includegraphics[width=\textwidth]{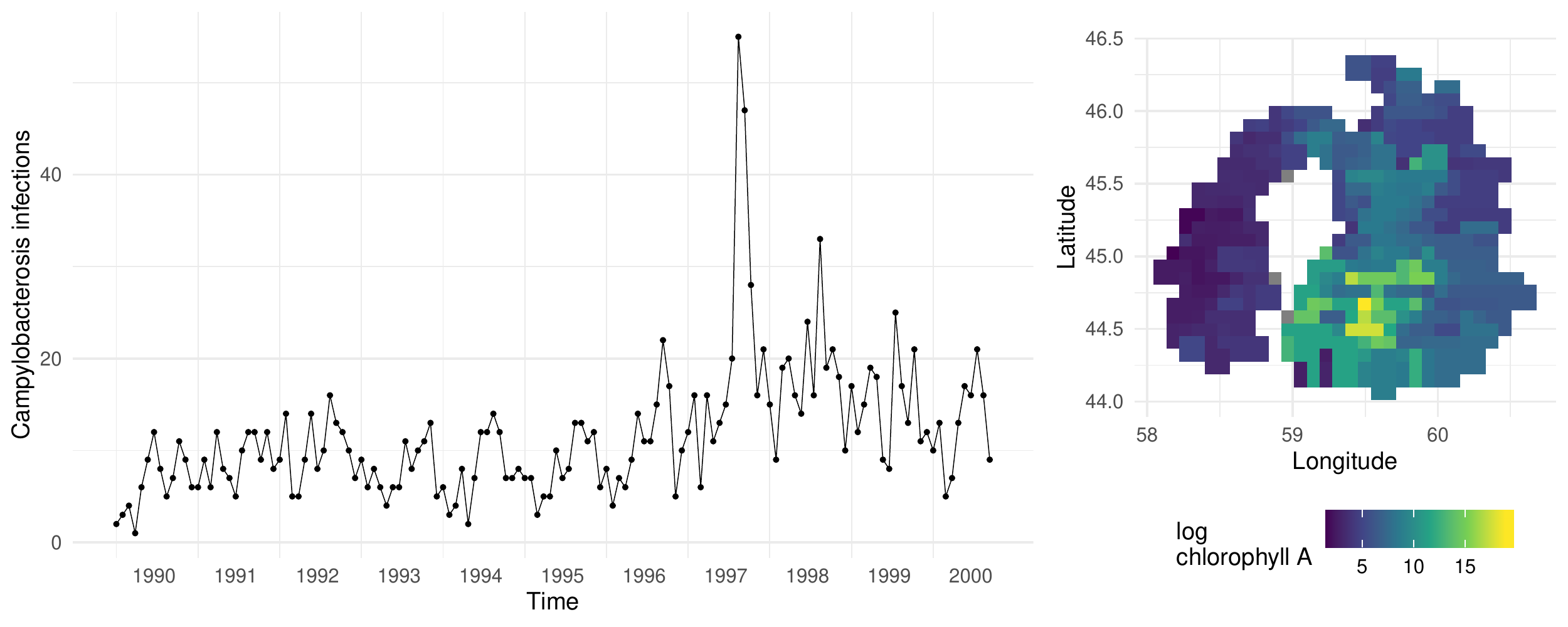}
\caption{Examples of data with underlying dependence between observations. Left shows counts of campylobacterosis infections in northern Qu\'ebec, summarized every 28 days from 1 January 1990 to 31 October 2000. Right shows the raw $\log$ chlorophyll A in the Aral sea from the SeaWIFS satellite. In both cases we can build a model that takes into account the structure in the data.}
\label{fig:exdat}
\end{figure}

An example of this kind of data might be a time series of counts. The left panel of Figure \ref{fig:exdat} show human cases of campylobacterosis (a common form of food poisoning, often originating in under-cooked poultry) in northern Qu\'ebec, every 28 days from 1 January 1990 to 31 October 2000. We may expect a given time period's count to be similar to its neighbours (e.g. due to seasonal variation), so our aim is to build a model that can capture this dependence. Using the above formulation, we model the counts as Poisson \(z(x) \sim \text{Po}(\exp(f(x)))\) where $x$ represents time, $z(x)$ is the number of cases at time $x$, $f$ is a function of time representing the underlying process and $\exp$ is the appropriate inverse link function. Dependence structures become more complex when we move to a spatial domain. The right panel of Figure \ref{fig:exdat} shows remotely-sensed $\log$ chlorophyll A levels in the Aral sea, derived from satellite data. In this case we expect that pixels close to each other have similar chlorophyll A levels. We now have $x$ represent a location in space, and the $\log$ chlorophyll A level at that location, $z(x)$ is modelled by  \(z(x) = f(x) + \epsilon(x)\), where now $f$ is a 2-dimensional stochastic process and \(\epsilon(x) \sim N(0, \sigma^2)\). We revisit these examples in Section \ref{s:examples}, below.

Including \(f\) in the model raises two issues: how to specify the covariance function \(c\) and, once specified, how to fit the model. There are many possible solutions to these questions, including the SPDE and smoothing penalty approaches, and each uses different theoretical and numerical approximations; however, there is a common element: for observation locations \(\{x_1, \ldots, x_n\}\),  each method aims to define the covariance between these locations, by constructing an approximation to the precision matrix, defined as the inverse of the covariance matrix (\(\bQ = \bSigma^{-1}\)). The precision matrix is a fundamental quantity required to fit these models \citep{simpson_order_2012}. 
The size of \(\bSigma\) and \(\bQ\) makes the necessary computations expensive, in particular, if one of these matrices needs to be inverted so as to compute the other. It is in trying to avoid this computational burden that approximations are used. 

The SPDE approach provides a method to approximate \(\bQ\) without the common substantial computational burden. The SPDE is an equation to be solved. Solutions to this equation are stochastic processes whose covariance structure is chosen to satisfy the relationship the SPDE specifies. The SPDE approach involves finding an SPDE whose solutions have the covariance structure, and implied precision matrix, that is desired for \(f\). \citet{lindgren_explicit_2011} show how to find an approximate solution to the SPDE by representing \(f\) as a sum of basis functions multiplied by coefficients; this provides a computationally efficient way to compute \(\bQ\): \citet{lindgren_explicit_2011} show that the coefficients of these basis functions form a Gaussian Markov random field, for which methods for fast computation of precision matrices already exist \citep{rue_gaussian_2005}. These computations make it possible to fit models quickly using integrated nested Laplace approximations \citep[INLA;][]{rue_approximate_2009}.

\citet{rue_review_2017} is a comprehensive review of INLA that defines the class of latent Gaussian models with additive linear predictors which INLA is designed to fit.  They also introduce Gaussian Markov random fields and their properties that lead to efficient computation, a key feature of the SPDE approach.   
\citet{bakka_review_2018} review the use of INLA for spatial modelling with a focus on the SPDE approach.  They give an intuition for the method by introducing the notion of a ``discretised" differential operator and describing the finite element methods that are used to solve the SPDE \citep{brenner_mathematical_2007, bakka_solve_2018}. See also \citet{krainski_advanced_2019} for a collection of worked examples of modelling with SPDEs using \texttt{R-INLA}, \citet{wood_simplified_2019} for an approach to nested Laplace approximations without sparse Gaussian Markov random field structures, and \citet{blangiardo_spatial_2015} for a comprehensive textbook on spatio-temporal modelling with \texttt{R-INLA}.  
We note that the \texttt{R-INLA} implementation of the SPDE approach has been applied in a wide variety of domains such as spatial epidemiology \citep{arab_epidemiological_2015}, species distribution mapping \citep{derivera_species2018}, spatial point processes \citep{simpson_going_2016, yuan_point_2017, soriano-redondo_spatio-temporal_2019} and environmental science \citep{huang_evaluating_2017}, to name just a few examples.  
Our presentation here differs from the above  resources in that we explicitly draw links with another well-known modelling framework. 

The basis-penalty smoothing approach \citep{wood_generalized_2017} is similar to the SPDE approach: the function \(f\) is a sum of basis functions multiplied by coefficients
. Rather than specify an SPDE and deduce a covariance structure between the coefficients, a smoothing penalty is used to induce correlation between the coefficients. This penalty measures how smooth \(f\) is in its domain; intuitively, if \(f\) changes more smoothly then values of \(f\) at nearby locations are more correlated. Jointly optimising a measure of fit (sum of squares or log-likelihood) and smoothing penalty leads to an optimal curve, the smoothing spline \citep{wahba_spline_1990}.  This is a well-established approach with several excellent introductory resources \citep{hastie_generalized_1990, ramsay_functional_2005, wood_generalized_2017} and has been applied in many spatio-temporal modelling contexts (recent examples include  \citet{wood_smoke_2017, simpson_modelling_2018, pedersen_hierarchical_2019})    

There is a direct correspondence between smoothing splines and stochastic processes \citep{kimeldorf_correspondence_1970}: the smoothing spline is a minimum variance unbiased linear estimator of the posterior mean of the stochastic process. For a stochastic process with a given covariance function, there is a corresponding SPDE and smoothing penalty such that one can estimate the posterior mean of \(f\) using the SPDE approach or the basis-penalty approach: both methods estimate the same quantity with the only differences being in numerical approximations and terminology. This means that the SPDE can be interpreted as a smoothing penalty and vice-versa.  

This equivalence has been confirmed by \citet{fahrmeir_bayesian_2001}, \citet{lindgren_second_2008}, and \citet{yue_bayesian_2014} who show how basis-penalty smoothers in a Bayesian framework can be interpreted within the SPDE paradigm. \citet{simpson_order_2012} remark that the SPDE formulation is useful because it provides those with a background in physics or applied mathematics a way to understand and apply the model. In contrast, less emphasis has been placed on discussing this equivalence the other way around: SPDE methods can be formulated as basis-penalty smoothers. The SPDE formulation can seem opaque and fundamentally different for those unfamiliar with the mathematical concepts used. For this reason, showing the approach within the familiar smoothing framework demystifies the workings of the model and allows researchers in other fields to understand, adapt, and use the methods.  
We note that our approach is aligned with the general aim of emphasising links between Gaussian processes and the reproducing kernel Hilbert spaces theory that underpins the basis-penalty smoothing approach \citep{kanagawa_gaussian_2018}, although here we take a more applied perspective.  

Our aim in this paper is to show that the SPDE model as introduced by \citet{lindgren_explicit_2011} (usually fitted using \texttt{R-INLA}) can be described as a basis-penalty smoother and fitted using \texttt{mgcv}. To do this, we first describe the SPDE method for those unfamiliar with the mathematical concepts used, highlighting the key steps in the method. Afterward, we show the equivalences and differences between the SPDE method and the analogous basis-penalty smoother. 

\section{The SPDE approach}
\label{s:spde}

\subsection{What is an SPDE?} 

A stochastic partial differential equation involves stochastic processes and differential operators. Examples of differential operators (\(D\)) are the first derivative, the second derivative, the gradient operator in two-dimensions or the Laplacian in two dimensions. Combinations of these are also differential operators, e.g., \(D = \mathrm{d}/\mathrm{d}x + \mathrm{d}^2/\mathrm{d}x^2\) such that \(Df = \mathrm{d}f/\mathrm{d}x + \mathrm{d}^2f/\mathrm{d}x^2\) for a function \(f\).  
Here, we consider only linear differential operators, that is, \(Df\) is a linear combination of derivatives of \(f\), of different orders. Differential operators of stochastic processes can be treated similarly to those applied to ordinary functions, there is one key difference that we will highlight below. Overall, an SPDE states that the differential of a function \(f\) is equal to some known stochastic process, most commonly the white noise process, \(\epsilon\). The white noise process is completely uncorrelated and \(\epsilon(x)\) is a Normal random variable with mean zero and finite variance for every \(x\). 

In general, the SPDE states that \(Df = \epsilon\) for some differential operator \(D\). A stochastic process, \(f\), is called a solution to the SPDE if it satisfies this equation. Consider an example, let \(D\) be the first derivative of the function. The SPDE \(Df = \epsilon\) therefore states that the first derivative of \(f\) has mean zero and finite variance at every point; furthermore, it states that the value of the derivative at points \(x\) and \(y\) are uncorrelated for all \(x \neq y\). Approximately, this means that for a small \(\delta\) and point \(x\), \(f(x + \delta) = f(x) + \xi\) where \(\xi\) is a Normal random variable. Consider if the SPDE has a parameter \(\tau\) such that \(Df = \epsilon/\tau\) such that \(\tau\) controls the variance in the white noise process. This means that changes in \(f\) are more variable when \(\tau\) is reduced and less variable for higher \(\tau\). In other words, the parameters of the SPDE control the smoothness of \(f\). It is important to note that here the term ``smoothness" is not used in a mathematical way, meaning differentiability, nor in a strictly statistical way, referring to correlation range, but in a qualitative way---when we speak of differentiability or correlation we shall use those terms explicitly.  

For a given \(D\), the mathematical form of the solution to the SPDE \(Df = \epsilon\) is known: \(f(x) = \int w(x - u)\epsilon(u)\;\mathrm{d}u\) where \(w\) is a function you can derive given you know \(D\). The function \(w\) is called Green's function; in the appendix (Proposition $1$) we show how this function is derived from \(D\). Intuitively, \(w\) acts as a weighting function such that the value of the stochastic process at \(x\) is a weighted sum over the white noise process; this is called a convolution. Suppose \(w\) were set to give infinite weight to distance \(0\), \(w(0) = \infty\), and zero elsewhere, \(w(d) = 0\) for \(d \neq 0\), then \(f(x) = \epsilon(x)\): \(f\) is just the white noise process, completely uncorrelated. Alternatively, if \(w\) gave equal weight to all distances, e.g., \(w(d) = 1\) for all \(d\), then \(f(x)\) would be constant, perfectly correlated. Between these two extremes are weighting functions that reproduce correlations over different ranges. It can be shown that the covariance function is given by \(c(x,y) = \int w(x - u)w(y - u)\;\mathrm{d}u\), see appendix (Proposition $2$) for the derivation. 

In summary, the solutions to the SPDE \(Df = \epsilon\) have a covariance structure that is induced by the choice of \(D\). This means that one could describe a system using an SPDE and then deduce the associated covariance function from it. The power of the SPDE approach is realised by doing the opposite: find a \(D\) that induces the covariance function that you want. The power of finding the SPDE corresponding to a desired covariance function is that the precision matrix can be efficiently computed using the SPDE.

\subsection{Solving the SPDE} 

The SPDE involves applying a differential operator \(D\) to a stochastic process, \(f\), but this cannot be done in the same way as when you apply \(D\) to a known function. This is because \(f\) is random and, in many cases, realisations of \(f\) will not be suitably differentiable. For example, the Brownian motion stochastic process has a derivative equal to the white noise process, but it is also known that simulated trajectories of Brownian motion are nowhere differentiable.  \(Df = \epsilon\) is a convenient shorthand way to think about the SPDE, but technically, the SPDE only has meaning when stated in an integral form. That is, \(Df = \epsilon\) means that we require \(\int Df(x)\phi(x)\;\mathrm{d}x = \int \epsilon(x)\phi(x)\;\mathrm{d}x\) for every function \(\phi\) with compact support. The function \(\phi\) is often called the test function. 
For brevity, let \(\langle f, g \rangle = \int f(x)g(x)\;\mathrm{d}x\) and so the integral form is \(\langle Df, \phi \rangle = \langle \epsilon, \phi \rangle\). The notation \(\langle f, g \rangle\) is called the inner product of \(f, g\), it has many nice mathematical properties, including being linear, that is \(\langle \sum_{i=1}^n a_if_i, \sum_{j = 1}^m b_ig_i \rangle = \sum_{i=1}^n\sum_{j=1}^m a_ib_j\langle f_i, g_j \rangle\) for functions \(f_1,\ldots,f_n, g_1, \ldots, g_m\) and constants \(a_1,\ldots,a_n, b_1, \ldots, b_m\). 

In the integral form, the equation makes sense because any stochastic process can be integrated, but not every one can be differentiated. 
By requiring the equation to hold for every \(\phi\), we require the left-hand stochastic process \(Df\) and the right-hand process \(\epsilon\) to have the same integral, no matter how we average over space. For example, if the stochastic processes were one-dimensional, we could split the real line into intervals \([n,n+1]\) and select a function \(\phi_n\) to be one on this interval and zero outside. Since the integral equation must hold for all such functions, we therefore require \(Df\) to have the same average value as \(\epsilon\) on each and every interval. 

Given an SPDE, \citet{lindgren_explicit_2011} show how to derive an approximate solution using the finite element method. The domain (e.g., time or space) is split into ``elements", e.g., a grid or a triangulation, often called a mesh. To each point \(j = 1, \ldots, M\) on this mesh, a basis function  \(\psi_j\) is associated. The solution to the SPDE is then a weighted sum of the basis functions and random variables \(\beta_j\): \(f(x) = \sum_{j=1}^M \beta_j \psi_j(x)\).

The integral form of the SPDE then implies that for any function \(\phi\), \(\sum_{j=1}^M \beta_j \langle D\psi_j, \phi \rangle = \langle \epsilon, \phi \rangle\). We cannot, however, check this equation for infinitely many test functions \(\phi\), so instead we restrict to only testing with the functions that can be written in our chosen basis. As $D$ is a linear operator, this is equivalent to solving the system of equations \(\sum_{j=1}^M \beta_j \langle D\psi_j, \psi_i \rangle = \langle \epsilon, \psi_i \rangle\) for every \(i = 1, \ldots, M\). This system can be written as a matrix equation: \(\bP\bbeta = \bee\) where \(\bP\) has \((i,j)^{\text{th}}\) entry \(\langle D\psi_i, \psi_j \rangle\) and \(\bee\) has \(j^{\text{th}}\) entry \(\langle \epsilon, \psi_j \rangle\). 

To summarise, the SPDE is written in an integral form, sometimes using inner products, since stochastic processes are well defined when integrated but not when differentiated. Given this, the solution is represented in a chosen basis. The integral form is then solved by considering only test functions within that basis. This leads to a matrix equation involving the coefficients \(\bbeta\), the matrix \(\bP\), and the random vector \(\bee\). The random vector \(\bee\) has known distribution, because it depends only on the basis functions and the white noise process: \(\bee\) has a multivariate Gaussian distribution with mean zero and a precision matrix \(\bQ_e\) where \(\bQ_e^{-1}\) has \((i,j)^{\text{th}}\) entry \(\langle \psi_i, \psi_j \rangle\). It follows from \(\bP\bbeta = \bee\) that 
\(\bbeta \sim N(0, \bQ^{-1})\) where \(\bQ = \bP^T\bQ_e\bP\). The SPDE is therefore a way to specify a prior for \(\bbeta\).   

This provides an approximate solution to the SPDE. For example, given an SPDE, one can use the finite element method to compute \(\bQ\) and therefore simulate \(\tilde{\bbeta}\) from a multivariate Gaussian distribution with precision \(\bQ\). The function \(\tilde{f} = \sum_{j = 1}^M \tilde{\beta}_j \psi_j\) would then be a realisation from a stochastic process which is a solution to the SPDE, a stochastic process with the covariance structure implied by \(D\).

\subsection{Mat\'ern SPDE} 

\begin{figure}[H]
\centering
\includegraphics[width=\textwidth]{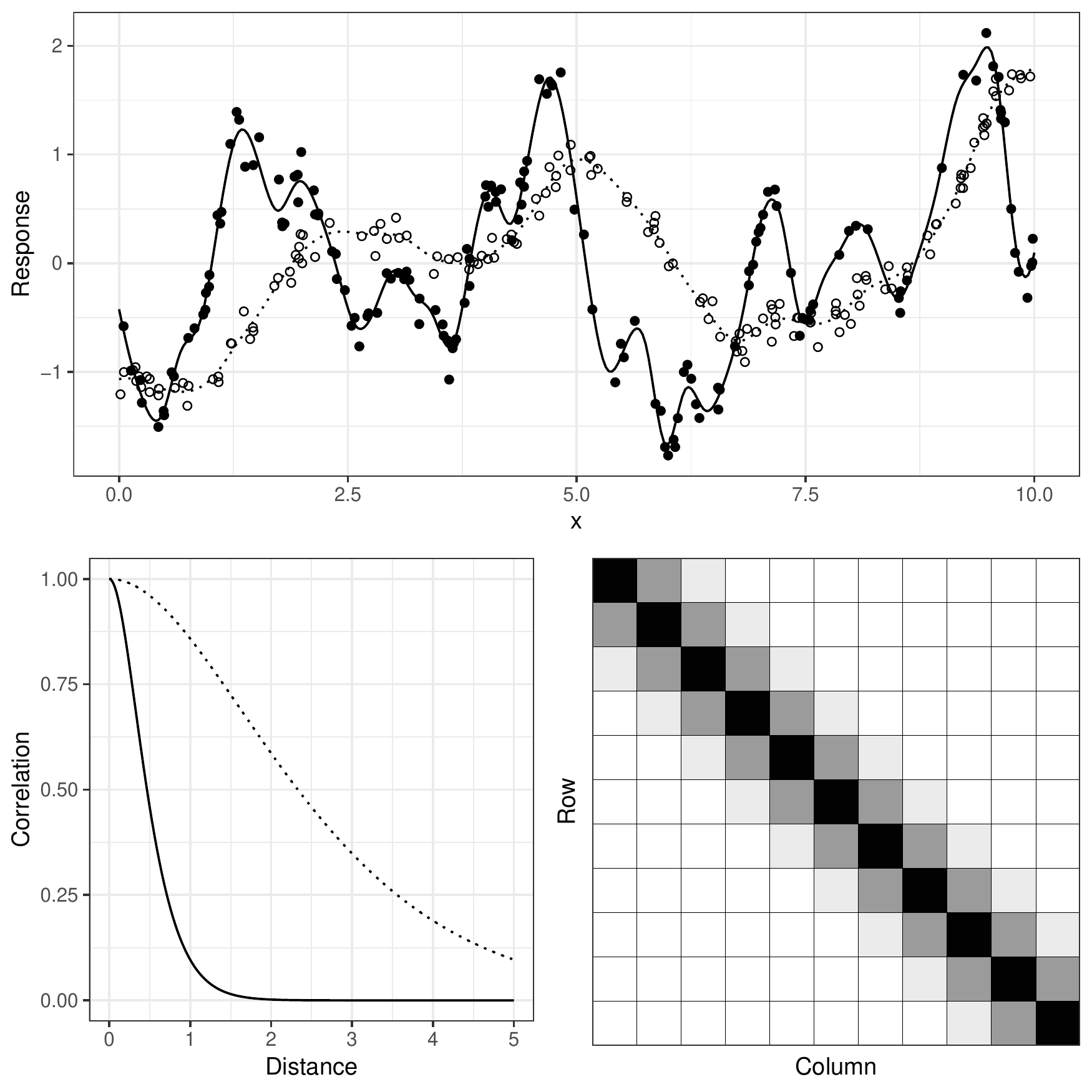}
\caption{Two functions, one smooth (long-range correlation, dashed line, open circle data) and one rough (short-range correlation, solid line, filled circle data) (top plot), their Mat\'ern correlation functions (bottom left plot, same line types) and the first 11 rows and columns of an example approximate Gaussian Markov Random field precision matrix (bottom right plot, darker shade indicates higher absolute value, each row and column corresponds to a data point location).}
\label{fig:smoocor}
\end{figure}

The focus of \citet{lindgren_explicit_2011} and the covariance function most commonly used in the \texttt{R-INLA} software is the Mat\'ern covariance function. The Mat\'ern covariance function is considered a flexible model for the dependencies found in real world observations: it has the form \( c(x, y) = \dfrac{2^{1-\nu}}{(4\pi)^{d/2}\kappa^{2\nu}\tau^2\Gamma(\nu + d/2)} (\kappa \|x - y\|)^{\nu}K_{\nu}(\kappa\|x - y\|) \) where \(\nu, \kappa, \tau\) are parameters, \(K_{\nu}\) is the modified Bessel function of the second kind, and $d$ is the dimension of the domain. Figure \ref{fig:smoocor} shows realisations from two stochastic processes with Mat\'ern covariance functions in one-dimension, one with a longer correlation range than the other. 

It is difficult to fit models with this covariance structure due to the computational issues mentioned above. \citet{lindgren_explicit_2011} apply the finite element method to approximate stochastic processes with Mat\'ern covariance (a comparison of the notation used in Section 2.2 and that used in \citet{lindgren_explicit_2011} is given in the appendix, Section 5). To do this, they present the differential operator that corresponds to this covariance function: \(D = (\kappa^2 - \Delta)^{\alpha/2}\tau\) where \(\alpha = \nu + d/2\). 

When \(\alpha \neq 2\), this is called a fractional differential operator; for this paper, we consider only the case when \(\alpha = 2\) and so \(D\) is again a linear differential operator. In practice, \(\alpha\) is poorly identified and difficult to estimate from data, so its value is often assumed to be fixed \citep{zhang_inconsistent_2004}. 

\citet{lindgren_explicit_2011} solve the SPDE \(\kappa^2f - \Delta f = \epsilon/\tau\) using the finite element method. By deriving the weighting function and computing the covariance from this SPDE, \cite{whittle_stationary_1954} shows that the solutions have Mat\'ern covariance, as desired. In other words, the precision matrix computed from the finite element method is an approximation to the precision matrix one would obtain if you computed the variance-covariance matrix \(\bSigma\) with the Mat\'ern covariance function and then, at great computational cost, inverted this matrix. Figure \ref{fig:smoocor} shows a subview of the approximate precision matrix: the matrix is mostly filled with zeroes, with non-zero values occurring on three bands down the diagonal. This is an example of a sparse matrix, computations with these matrices are efficient because many of the computations ordinarily required can be omitted as it is known the matrix is mostly zeroes. 

To use the finite element method one must chose a mesh, a grid or triangulation, over the domain and a basis to define on this mesh. The default choice in \texttt{R-INLA} is to use a regular grid in 1D (or a constrained Delaunay triangulation in 2D) to produce a mesh and then define piecewise linear basis functions (specifically, linear B-spline basis functions) on this mesh.

\section{The basis-penalty approach}
\label{s:smoothing}

\subsection{What is a basis-penalty smoother?} 

The basis-penalty approach refers to models where \(f\) is assumed to have the form \(f(x) = \sum_{j = 1}^M \beta_j\psi_j\) for \(M\) basis functions \(\psi_1, \ldots, \psi_M\) and parameters \(\bbeta = (\beta_1, \ldots, \beta_M)\). A model is then assumed for the observations given this form for \(f\) to provide a measure of fit, the log-likelihood \(l(\bbeta)\). Alternatively, the sum-of-squares can be used as a measure of fit. Optimising to obtain \(\hat{\bbeta}\) leads to a function \(\hat{f}\)
that, given \(M\) is large enough, will interpolate the observed data: capturing the noise in the observations as well as the underlying signal (such overfitting stops us from making inference on the signal). A smoothing penalty, \(J(\bbeta, \lambda)\), is subtracted from the log-likelihood to penalize functions that are too wiggly. The smoothing parameter, \(\lambda\), controls the extent of the penalization (a larger value of $\lambda$ leads to a smoother $\hat{f}$).

The estimates for \(\bbeta\) are defined to be those that optimise the joint measure of fit and smoothness, the penalised likelihood: \(l_p(\bbeta, \lambda) = l(\bbeta) - J(\bbeta, \lambda)\). This involves estimating both the optimal smoothing parameter \(\lambda\) and coefficients \(\bbeta\). In practice, REstricted Maximum Likelihood \citep[REML;][]{wood_fast_2011} is used to to do this. 

There are several choices for the smoothing penalty. Most are defined using a differential operator \(D\).
For example, in one dimension, the smoothing penalty \(J(\bbeta, \lambda) = \lambda \int (\partial^2 f / \partial x^2)^2 \;\mathrm{d}x\), i.e., where \(D\) is the second derivative, is often used. For this penalty, functions with rapidly changing gradients are penalised while functions with constant gradient, straight lines, have no penalty. In higher dimensions, the thin-plate spline \citep{wood_thin_2003} is often used with penalty: \(J(\bbeta, \lambda) = \lambda \int \left(\partial^2 f / \partial x^2\right)^2 + 2\left(\partial^2 f / \partial x \partial y\right)^2 + \left(\partial^2 f / \partial y^2\right)^2 \mathrm{d}x\mathrm{d}y\) for two-dimensions. This penalty takes the total variation in the gradient of \(f\) including the interaction between the coordinates. The penalty for smoothing splines takes the form \(J(\bbeta, \lambda) = \lambda \int (Df)^2 \;\mathrm{d}x\) for some chosen differential operator \(D\) (see \citet{yue_stationary_2010} and \citet{yue_bayesian_2014} who show this for the thin plate spline penalty). This can also be written as an inner product \(J(\bbeta, \lambda) = \lambda \langle Df, Df \rangle\). 

When \(f(x) = \sum_{j = 1}^M \beta_j\psi_j(x)\), the penalty based on the differential operator \(D\) can be written in matrix form: \(J(\bbeta, \lambda) = \lambda\bbeta^\T \bS \bbeta\) where \(\bS\) is a \(M \times M\) matrix with \((i,j)^{\text{th}}\) entry \(\langle D\psi_i, D\psi_j \rangle\).

In summary, a basis-penalty smoother is specified by selecting a basis, e.g., a B-spline basis of specified order, and a smoothing penalty. The parameters are then estimated by optimising the penalised likelihood: \(l_p(\bbeta, \lambda) = l(\bbeta) - \lambda \bbeta^\T\bS\bbeta\).

\subsection{Connection between SPDE and penalty} 

Rewriting the penalized log-likelihood as a likelihood we obtain \(\exp\{l_p(\bbeta, \lambda)\} = \exp\{l(\bbeta)\}\times\exp(-\lambda\bbeta^\T \bS \bbeta)\). A Bayesian interpretation of the penalised likelihood  as proportional to a posterior implies that \(\exp(-\lambda\bbeta^\T \bS \bbeta)\) is an improper prior for \(\bbeta\) \citep{silverman_aspects_1985, wood_generalized_2017}. Since \(\exp(-\lambda\bbeta^\T \bS \bbeta)\) is proportional to a multivariate normal distribution with mean zero and precision matrix \(\bS_{\lambda} = \lambda\bS\), the penalized likelihood is equivalent to assigning the prior \(\bbeta \sim N(0, \bS_{\lambda}^{-1})\). 

The connection between the SPDE approach and the basis-penalty approach can now be made clear. It can be shown that for a given differential operator \(D\), the approximate precision matrix for the SPDE \(Df = \epsilon\) is the same as the precision matrix \(S_{\lambda}\) computed using the smoothing penalty \(\langle Df, Df \rangle\) (appendix, Proposition 3).

This connection has two implications. First, it means that the differences between the basis-penalty approach and the SPDE finite element approximation, when using the same basis and differential operator, are differences in implementation only, as both should lead to the same approximate precision matrix. Second, the connection means that any SPDE of the form \(Df = \epsilon\) can be understood and interpreted as a smoothing penalty of the form \(\langle Df, Df \rangle = \int \{Df(x)\}^2\;\mathrm{d}x\), and vice-versa.

\subsection{Mat\'ern penalty}

The SPDE specified in \citet{lindgren_explicit_2011} has the differential operator \(D = \tau(\kappa^2 - \Delta)\). Given the connection described above, this can be interpreted as a smoothing penalty: \(\tau^2 \int (\kappa^2 f - \Delta f)^2 \;\mathrm{d}x\). This penalty is different from those considered above because it contains two smoothing parameters: \(\tau\) and \(\kappa\). This offers it more flexibility. The penalty can still, however, be interpreted as such: it is a trade-off between the value of the function \(f\) and the second derivative \(\Delta f\) in each direction. As \(\kappa\) is increased, the value of \(\kappa^2 f\) increases, meaning that \(\Delta f\) can be higher, the function be less smooth, while keeping the penalty the same. Alternatively, \(\kappa\) can be described as the inverse correlation range: higher values of \(\kappa\) lead to less smooth functions meaning values of the function become less correlated. The smoothing parameter \(\tau\) controls the overall smoothness of \(f\). 

The Mat\'ern penalty can be written in matrix form as above, but for computational convenience, it is first broken into three parts: \(\langle Df, Df \rangle = \tau^2 (\kappa^4 \langle f, f \rangle + 2\kappa^2 \langle \nabla f, \nabla f \rangle + \langle \Delta f, \Delta f \rangle)\). Notice that it appears that the Laplacian \(\Delta\) has been replaced with the gradient operator \(\nabla\): this relationship holds here using Green's first identity and the Neumann boundary condition, see \citet{bakka_review_2018} for more detail. This leads to the smoothing matrix \(\bS = \tau^2 (\kappa^4 \boldsymbol{C} + 2\kappa^2 \boldsymbol{G}_1 + \boldsymbol{G}_2)\) where \(\boldsymbol{C}, \boldsymbol{G}_1, \boldsymbol{G}_2\) are all \(M \times M\) matrices with \((i,j)^{\text{th}}\) entries \(\langle \psi_i, \psi_j \rangle\), \(\langle \nabla \psi_i, \nabla\psi_j \rangle\), and \(\langle \Delta \psi_i, \Delta \psi_j\rangle\), respectively. All of these matrices are sparse and so computation of the smoothing penalty, \(\bbeta^\T\bS\bbeta\), is computationally efficient. The matrix \(\bS\) is equal to the matrix \(\bQ = \bP^\T\bQ_e\bP\) computed using the finite element method (Appendix, Proposition 3).

\subsection{Fitting the Mat\'ern SPDE in \texttt{mgcv}}
\label{spde-in-mgcv}

The \texttt{mgcv} R package allows the specification of new basis-penalty smoothers by writing new ``\texttt{smooth.construct}'' functions which build an appropriate design matrix (containing evaluations of the basis functions), penalty matrices and other optional components. Within this framework we can fit the SPDE model in \texttt{mgcv} providing a \texttt{smooth.construct.spde.smooth.spec} constructor. \texttt{R-INLA} provides helper functions to construct the required design and penalty matrices. Here we sketch an algorithm for setting-up SPDE models in \texttt{mgcv}. 

Given we have a response $\left\{y_i; i=1,\ldots, n\right\}$ and covariates in an $n \times n_c$ matrix $\bX_c$, we construct our model as follows.

\begin{enumerate}
  \item Create a mesh using \texttt{INLA::inla.mesh.1d} or \texttt{INLA::inla.mesh.2d}.
  \item Calculate $\mathbf{C}$,  $\mathbf{G}_1$ and $\mathbf{G}_2$ using \texttt{INLA::inla.mesh.fem} (\texttt{c1}, \texttt{g1} and \texttt{g2}, respectively).
  \item We need to connect the basis representation of \(f\) to the observation locations. At present \(\bbeta\) contains the value of \(f\) at each mesh point, not at each observation location. A matrix multiplication is used to project the values at all mesh points to the observations locations, it is called the projection matrix \(\mathbf{A}\) (found using \texttt{INLA::inla.spde.make.A}). The full design matrix \(\bm{X}\) is then given by combining the fixed effects design matrix \(\bX_c\) and the contribution for \(f\), \(\mathbf{A}\). 
  \item Having constructed the design matrix and penalty matrices, use REML to find optimal $\kappa$, $\tau$ and $\bbeta$ subject to the penalty matrix $\kappa^4\mathbf{C} + 2\kappa^2\mathbf{G}_1 + \mathbf{G}_2$. (Parametrisation for this model in \texttt{mgcv} is given in Supplementary Material section 4.)
\end{enumerate}

As REML is an empirical Bayes procedure, we expect point estimates for $\hat{\bbeta}$ to coincide for the procedure outlined above and \texttt{R-INLA}. A uniform prior is implied for  the smoothing parameters ($\tau$ or $\kappa$); \texttt{R-INLA} allows for similar estimation by just using the modes of the hyperparameters $\kappa$ and $\tau$ (the \texttt{int.strategy="eb"} option). Proper priors could be used if step (4), above, was replaced by an MCMC scheme.

\section{Examples}
\label{s:examples}

We now compare the SPDE and basis-penalty models applied to three example datasets. We fitted the SPDE Mat\'ern model in both \texttt{R-INLA} and \texttt{mgcv}. Code for these examples is available as supplementary material.

\subsection{Campylobacterosis cases in Qu\'ebec}

\cite{ferland_integer-valued_2006} analyse a time series of (human) cases of campylobacterosis in northern Qu\'ebec, with observations every 28 days from 1 January 1990 to 31 October 2000 (140 observations). We  modelled the number of infections as a function of time, using a Poisson response and a $\log$ link function. The model is fitted using three approaches ($i$) a Mat\'ern basis-penalty smoother with 50 degree 2 B-splines, fitted with \texttt{mgcv}; ($ii$) a Mat\'ern SPDE for \(f\) with a finite element basis of 50 degree 2 B-splines and penalized complexity priors \citep{simpson_penalising_2017} on smoothing parameters, fitted with \texttt{R-INLA}; ($iii$) a basis-penalty smoother with penalty equal to the integral of the squared second derivative, using 50 degree 2 B-splines fitted using \texttt{mgcv}. 

\begin{figure}[htbp]
\begin{center}
\includegraphics[width=\textwidth]{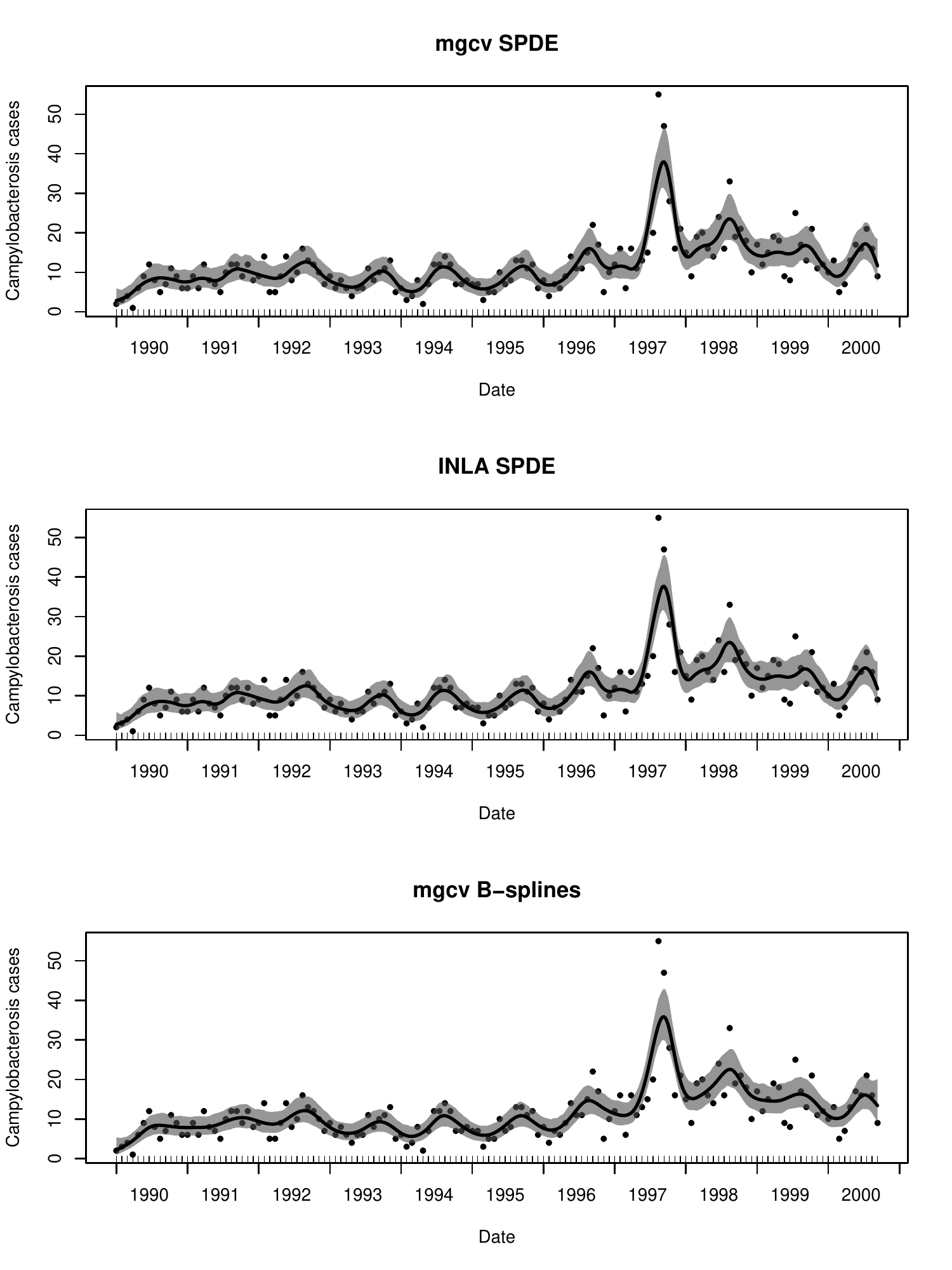}
\caption{Campylobacterosis cases modelled using: a Mat\'ern basis-penalty smoother fitted with \texttt{mgcv} (top), a Mat\'ern SPDE fitted with \texttt{R-INLA} (middle), a B-spline basis-penalty smoother fitted using \texttt{mgcv} (bottom).}
\label{fig:campy}
\end{center}
\end{figure}

Fitted models are shown in Figure \ref{fig:campy}. Results from the \texttt{R-INLA} and \texttt{mgcv} SPDE implementations are very similar. This is supported by the similarity in the estimated hyperparameters ($\tau=3.603$ and $\kappa=0.429$ for \texttt{R-INLA}, and $\tau=3.252$ and $\kappa=0.475$ for \texttt{mgcv}). The squared second derivative penalty B-spline fit from \texttt{mgcv} is smoother than those from the SPDE-based methods.

\subsection{Aral sea chlorophyll}

Moving to a 2-dimensional smoothing problem, we consider remotely sensed ($\log$) chlorophyll from the Aral sea collected by the NASA SeaWifs satellite over a series of 8 day observation periods. The 496 observations used here are averages (from 1998 to 2002) of the 38th observation period. Data were taken from the \texttt{gamair} package (dataset \texttt{aral}) and consist of spatial coordinates and logarithm of chlorophyll concentration.

We built a mesh using \texttt{fmesher::meshbuilder} (Supplementary Figure 2) and generated two-dimensional degree 1 B-splines. We consider the model \(y_i = f(\bm{x}_i) + \epsilon\) for location \(\bm{x}_i\) with no fixed effects. We fitted the Mat\'ern model using the SPDE and penalty approaches in \texttt{R-INLA} and \texttt{mgcv}. For the \texttt{R-INLA} model, penalized complexity priors were used. 

\begin{figure}[t]
\centering
\includegraphics[width=\textwidth]{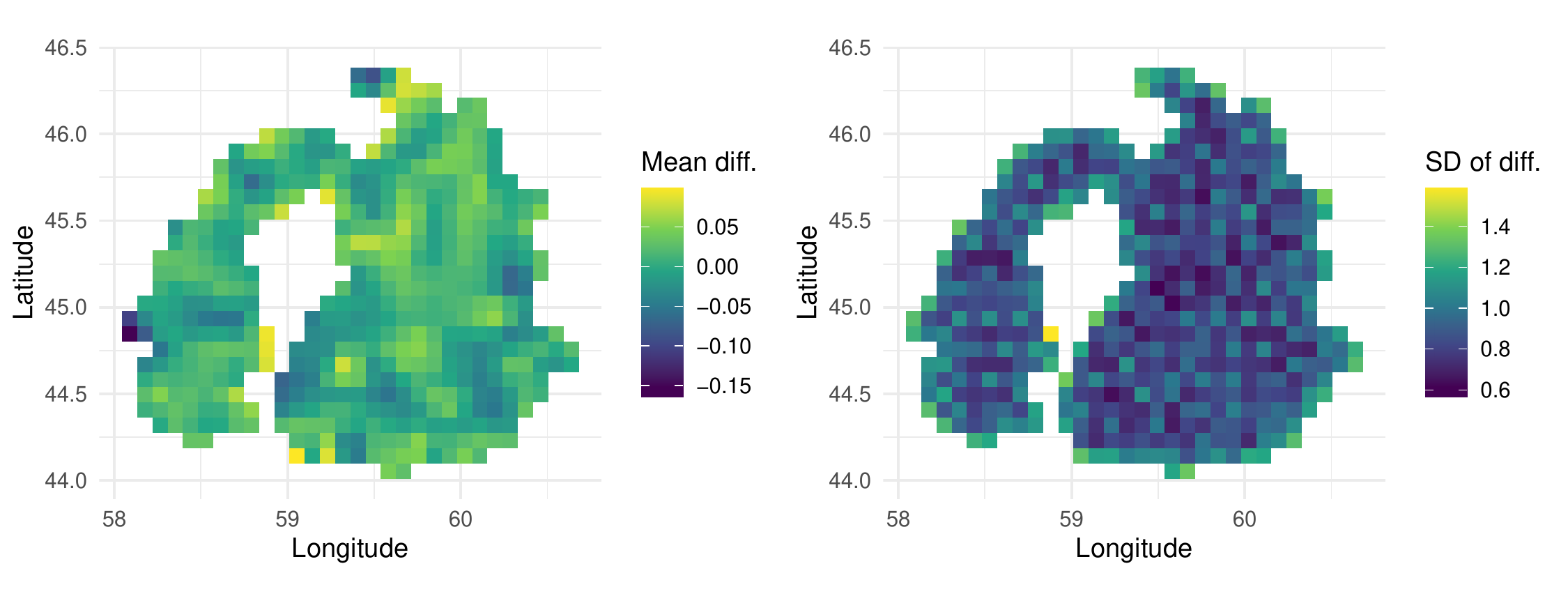}
\caption{Chlorophyll in the Aral sea example. Left shows mean difference in predictions and right shows standard deviation of the difference in predictions between SPDE models fitted using \texttt{mgcv} and \texttt{R-INLA}.}
\label{fig:aral}
\end{figure}

There was little visual difference with good agreement in the predictions (Supplementary Figure 3 shows the posterior mode and percentile credible surfaces for these models and Supplementary Figure 4 gives a graphical comparison). Hyperparameter estimates were similar: $\tau=0.059$ and $\kappa=3.43$ for \texttt{R-INLA} and $\tau= 0.059$ and $\kappa= 3.543$ for \texttt{mgcv}. To investigate differences between the two models we took (1000) samples from the posterior of each model and looked at the differences between pairs of realisations on a per-cell basis. Plots of the mean of these differences and their standard deviations are shown in Figure \ref{fig:aral}. The mean plot shows structure to the differences in the models, though differences are relatively small (range of $\log$ chlorophyll A values in original data: 1.905--19.275). This is to be expected if the models produce similar predicted values to each other, which are consistent through each realisation. 

\subsection{MODIS land surface temperatures}

To compare the two approaches on a large data set, we now turn to land surface temperature data collected by the Terra instrument on the MODIS satellite. The data consist of a  $500\times300$ grid of measurements in the latitude range 34.29519--37.06811 and longitude range -95.91153{--}-91.28381 on August 4, 2016. The training data (105,569 observations) as defined in \cite{heaton_case_2018} were used to fit the model, but a significantly simpler mesh was used (see Supplementary Figure 5). Following from \cite{heaton_case_2018} we assumed a Gaussian response for temperature and fitted a 2-dimensional SPDE model on latitude and longitude.

As the dataset was quite large, we used the \texttt{bam} (``big additive model'') function in \texttt{mgcv} to fit the SPDE model \citep[additionally discretizing covariate values for efficient storage and computation;][]{wood_smoke_2017}. The SPDE fitted by \texttt{R-INLA} used the empirical Bayes integration strategy. We timed the fitting for both approaches (ignoring mesh setup, which was shared across methods), taking only the time for \texttt{inla} and \texttt{bam} to run. The \texttt{mgcv} model was slightly faster (4.71 minutes versus 5.23 in \texttt{R-INLA} running on a Windows server using 1 core of a Xeon Gold 6152 at 2.1GHz with 512Gb RAM). Supplementary Figure 6 compares the predictions from the two methods, the largest absolute difference between predictions is 4.761, which is small compared to the range of the data.



\section{Discussion}
\label{s:discussion}

We have drawn links between two approaches to fitting the same model: the stochastic partial differential equation method as implemented in \texttt{R-INLA} and the basis-penalty smoothing approach as fitted in \texttt{mgcv}. This paper aims to make accessible what is equivalent between the approaches, what is a matter of choice, and what is fundamentally different. \cite{yue_bayesian_2014} show how splines can be specified using the SPDE approach, benefitting those familiar with SPDEs. Here, we do the opposite for the benefit of those familiar with the (penalized likelihood/empirical Bayes) GAM framework. 
Supplementary Figure 1 is a flow diagram showing the parallels between the smoothness and correlation approaches we have discussed.

Similarities between many smoothing techniques can be drawn. Smoothing splines, kriging, Gaussian Markov random fields, and SPDEs approximate similar models, but their explanations make it difficult for practitioners to appreciate their commonalities and determine precisely what is a necessary and what is a coincidental association. Taking the precision matrix  
as the common currency between these methods, 
a modelling framework emerges:
\begin{enumerate}
 \item \textbf{Choose a covariance model}: explicitly, as in kriging, through the smoothness penalty as with smoothing splines, or with an SPDE;
 \item \textbf{Approximate the precision matrix \(\bQ\)}: reduce dimension (fixed rank kriging, thin plate splines) or induce sparsity in \(\bQ\) (
 B-splines, SPDE);
 \item \textbf{Draw approximate inference using a software implementation}: e.g., with \texttt{mgcv}, MCMC (e.g., Stan \citep{carpenter_stan:_2017}; \texttt{JAGS}, \citep{plummer_jags_2017}), \texttt{R-INLA} \citep{rue_approximate_2009}, \texttt{lme4} \citep{bates_fitting_2015} or \texttt{TMB} \citep{kristensen_tmb_2016}.
\end{enumerate}

This paper is an example of comparing two methods according to this framework. Doing so for other smoothing methods will allow alternative modelling approaches to be compared on the grounds of genuine differences: in the covariance function, in the approximation for \(\bQ\), in the estimation procedure, or, simply, in the software implementation.

\section*{Acknowledgements}

The authors wish to thank the editor and two anonymous reviewers for their constructive comments on the first submission of the manuscript. They also thank Finn Lindgren for extremely helpful input and Simon Wood for the suggestion of the smoothing parameter parameterisation. The authors also thank Steve Buckland, David Borchers and Joe Watson for their comments on the manuscript.

\bibliographystyle{chicago}
\bibliography{paper}

\end{document}